\documentclass[journal]{IEEEtran}

\ifCLASSINFOpdf
\else
\fi

\usepackage{bm}

\usepackage{amsmath}
\usepackage{amsfonts}
\usepackage{mathtools}

\usepackage{comment}
\usepackage{booktabs}
\usepackage{multirow}
\usepackage[T1]{fontenc}

\begin{document}
\title{Regularized Three-dimensional Generative Adversarial Nets for Unsupervised Metal Artifact Reduction}

\author{Megumi~Nakao,~\IEEEmembership{Member,~IEEE,}
        Keiho~Imanishi,
        Nobuhiro~Ueda,
        Yuichiro~Imai,
        Tadaaki~Kirita
        and~Tetsuya~Matsuda,~\IEEEmembership{Member,~IEEE}
\thanks{M. Nakao and T. Matsuda are with the Graduate School of Informatics, Kyoto University, Yoshida-Honmachi, Sakyo, Kyoto 606-8501, JAPAN; e-mail: megumi@i.kyoto-u.ac.jp.}
\thanks{K. Imanishi is with e-Growth Co., Ltd., 403, Shimo-Maruya-cho, Nakagyo-ku, Kyoto 604-8006, JAPAN.}
\thanks{N. Ueda and T. Kirita are with the Department of Oral and Maxillofacial Surgery, Nara Medical University, Kashihara, Nara 634-0813, JAPAN.}
\thanks{Y. Imai is with the Department of Oral and Maxillofacial Surgery, Rakuwakai Otowa Hospital, Yamashina, Kyoto 607-8062, JAPAN.}

\thanks{Manuscript received Mar 31, 2020;} 
\thanks{We thank Stuart Jenkinson, PhD, from Edanz Group for editing a draft of this manuscript. This research was supported by JSPS Grant-in-Aid for Scientific Research (B) (grant number 19H04484). A part of this study was also supported by a JSPS Grant-in-Aid for challenging Exploratory Research (grant number 18K19918). }}

\markboth{Journal of \LaTeX\ Class Files, Mar~2020}%
{Nakao \MakeLowercase{\textit{et al.}}: Regularized 3D GANs for Unsupervised MAR}

\maketitle

\begin{abstract}
  The reduction of metal artifacts in computed tomography (CT) images, specifically for strong artifacts generated from multiple metal objects, is a challenging issue in medical imaging research. Although there have been some studies on supervised metal artifact reduction through the learning of synthesized artifacts, it is difficult for simulated artifacts to cover the complexity of the real physical phenomena that may be observed in X-ray propagation. In this paper, we introduce metal artifact reduction methods based on an unsupervised volume-to-volume translation learned from clinical CT images. We construct three-dimensional adversarial nets with a regularized loss function designed for metal artifacts from multiple dental fillings. The results of experiments using 915 CT volumes from real patients demonstrate that the proposed framework has an outstanding capacity to reduce strong artifacts and to recover underlying missing voxels, while preserving the anatomical features of soft tissues and tooth structures from the original images.
\end{abstract}

\begin{IEEEkeywords}
  Computed tomography, generative adversarial network, metal artifact reduction, unsupervised image translation
\end{IEEEkeywords}

\IEEEpeerreviewmaketitle

\section{Introduction}
\IEEEPARstart{M}{edical} procedures such as diagnosis, surgical planning, and radiotherapy can be seriously degraded by the presence of metal artifacts in computed tomography (CT) imaging. Metal objects such as dental fillings, fixation devices, and other electric instruments implanted in patients' bodies inhibit X-ray propagation \cite{Man99}, preventing accurate calculation of the CT values during image reconstruction and yielding dark bands or streak artifacts in the CT images \cite{Gjesteby16}\cite{Park16}. To correct the images, missing CT values for the underlying anatomical features must be compensated at the same time as the artifacts are removed. Although doctors make clinical efforts to manually collect such artifacts, this is a labor-intensive and time-consuming task. Many researchers have studied image filtering or reconstruction methods \cite{Gjesteby16}\cite{Meyer10}\cite{Zhang11}\cite{Mehranian13}\cite{Sunil16}, but metal artifact reduction (MAR) remains a challenging problem \cite{Huang15}\cite{Andersson18}\cite{Neroladaki19}, and no standard algorithm for strong, complex artifacts with missing pixels derived from multiple metal objects has yet been established.

The MAR methods commonly applied after image acquisition are filtering or normalization approaches in the projection domain \cite{Meyer10}\cite{Zhang11}\cite{Huang15}. Traditional image interpolation and iterative reconstruction approaches require physical models during CT scanning, and do not achieve sufficient artifact reduction against various shape and material characteristics of metals. In recent decades, statistical compensation techniques using prior knowledge of the artifacts have been investigated \cite{Stayman12}\cite{Wang13}\cite{Hegazy16}\cite{Haewon17}. The application of deep learning to medical images has gained significant interest, and has been actively studied in recent years \cite{Jin17}\cite{Chen17}\cite{Gjesteby17}\cite{Zhang17}. Supervised learning for artifact reduction requires an artifact-free image that corresponds to the image with artifacts; in practice, the preparation of such paired images is clinically difficult. Thus, sinograms or CT images generated with simulated typical metal artifacts are used as training data \cite{Zhang18}\cite{Huang18}. The usage of synthesized images enables high-quality image reconstruction under the condition that the three-dimensional shape and position of the metal are assumed to be known. However, this approach struggles to generate realistic artifacts that fully cover the complexity of real physical phenomena encountered during X-ray propagation, specifically in cases with multiple metal objects. Determining the variations of simulated artifacts from those of real ones in the CT images remains a challenge. 

Recently, generative adversarial networks (GANs) \cite{Goodfellow14} and their extensions \cite{Taigman16}\cite{Zhu17} have been extensively studied as a framework for unsupervised image-to-image translation. Unsupervised learning in the GAN framework does not require paired images, as a mapping function from the input to the target image domains is obtained by constructing a generator with the ability to transfer the image features. The generator is trained adversarially using a discriminator that attempts to distinguish whether the input is a synthesized image, leading to elaborate image-to-image translation. Extensive research on GAN-based medical image synthesis has been conducted for various clinical applications \cite{Yi19}. For instance, low-dose CT denoising \cite{Wolterink17}\cite{Yang18}, cross-modality transfer \cite{Nie18}\cite{Emami18}\cite{Liang19}\cite{Kida19}, and an application to organ segmentation \cite{Dai17} have been actively studied. 

The application of GANs to artifact reduction is a relatively new challenge, as technical difficulties mean that various low-quality images affected by strong metal artifacts exist in clinical CT images. Recent studies have applied GANs to MAR in small regions of CT images of the ear \cite{Wang18}\cite{Noble19}\cite{Wang19}. Du et al. \cite{Du18} presented preliminary results from GAN-based MAR for images with dental fillings, while Liao et al. \cite{Liao19} proposed a CycleGAN-based artifact disentanglement network and compared quantitative evaluation results against existing supervised/unsupervised MAR methods using synthesized datasets. However, the performance of GAN-based MAR remains to be clarified when the network is trained with clinical CT images containing complex artifacts derived from multiple dental fillings. Image correction in MAR should target artifact-affected regions and recover the underlying image features, while preserving the other regions with the native anatomical structures of the patients. To address these issues, we have focused on adversarial training of real patient datasets and the importance of learning three-dimensional (3D) features from the CT volumes. 

\begin{figure}[t]
  \begin{center}
  \includegraphics[width=85mm]{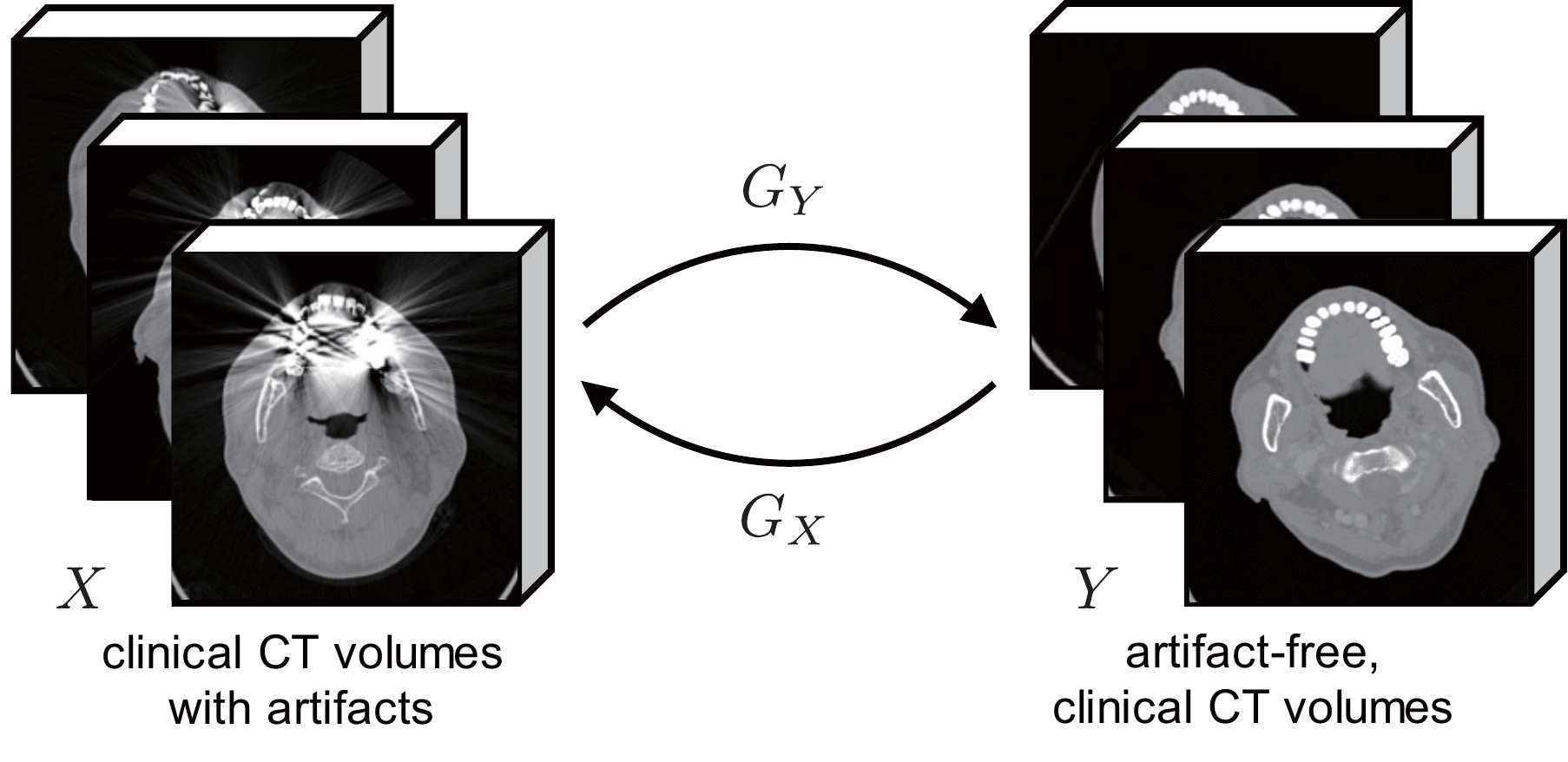}
  \caption{3D generative adversarial nets for unsupervised MAR of CT images, $X$: image domain with artifacts, $Y$: image domain without artifacts. The database is constructed from 915 clinical CT volumes consisting of head and neck images.}
  \label{fig:concept}
  \end{center}
\end{figure}

In this paper, we introduce MAR methods based on volume-to-volume translation learned from unpaired clinical CT images. The proposed methods are established based on an unsupervised learning scheme in the absence of synthesized images or simulated artifacts. 3D GANs are developed as an extension to the image-to-image translation framework of CycleGAN \cite{Zhu17}, and a mapping function for artifact reduction is trained using a patient CT volume database (see Fig. \ref{fig:concept}). The database is constructed from 915 clinical CT volumes: metal-free CT volumes and those with various patterns of metal artifacts derived from multiple dental fillings. There are infinitely many mappings that translate artifact-free to artifact-affected volumes, and we demonstrate that an adversarial objective with cycle consistency loss often fails to preserve anatomical features. We therefore seek a regularized loss function that captures the characteristics of metal artifacts, thus addressing the main issues faced by GAN-based MAR: recovering the underlying image features while preserving the native anatomical structures in artifact-free regions. We compare the proposed method against existing unsupervised approaches to clarify the MAR performance. Quantitative and qualitative evaluations show that the proposed framework outperforms the baselines for a wide range of clinical images with artifacts. Experiments involving the opinions of expert surgeons confirm the clinical applicability of the proposed 3D adversarial nets. 

The contributions of this study can be summarized as follows: 1) 3D generative adversarial nets are developed for unsupervised MAR, 2) a regularized loss function is designed for stable learning of the target features of artifacts, 3) a feasibility study of the volume-to-volume translation directly learned from clinical images is conducted, and 4) the quantitative performance of the proposed framework is evaluated and clinically validated with expert surgeons.

\section{Methods}
The goal of the proposed unsupervised learning method is to learn mapping functions between two domains: $X$, containing the CT volumes with artifacts, and $Y$, containing artifact-free CT volumes, given the training samples $\{x_i\} \in X$ and $\{y_j\} \in Y$. This section describes the details of our training datasets, 3D adversarial training scheme, the regularized mapping function, and the volume-to-volume translation process.  

\begin{figure*}[ht]
  \begin{center}
 \includegraphics[width=180mm]{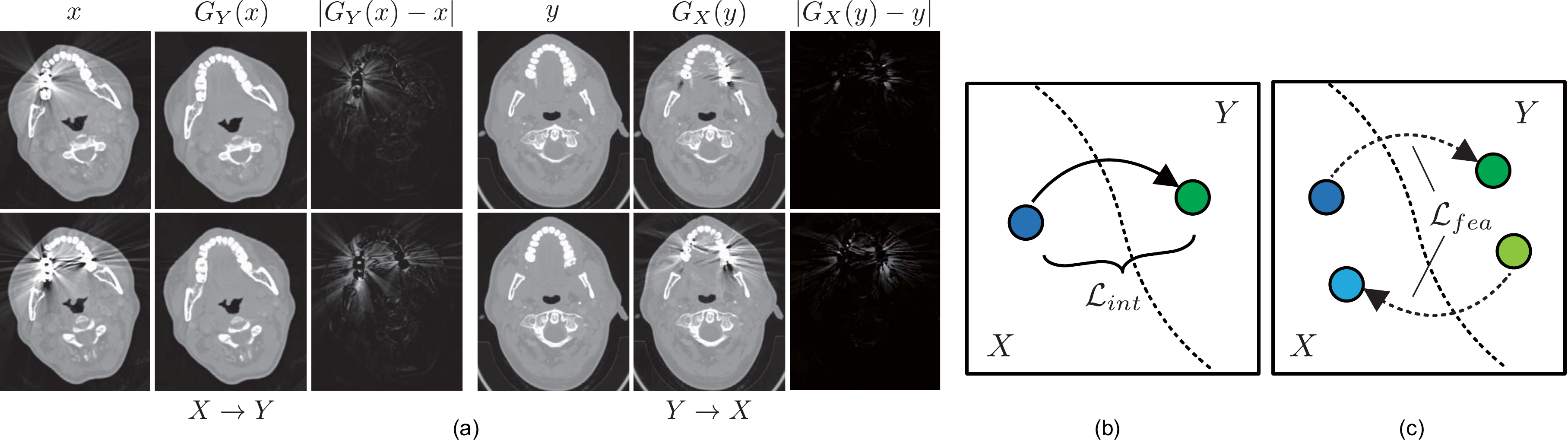}
  \caption{Adversarial training example and regularization of the mapping function: (a) two image slices $x$ and $y$ are translated to the other domain ($Y$ and $X$, respectively) based on the sparsity of artifacts, (b) intensity loss, and (c) artifact feature loss to induce the generators targeting the translation of metal artifacts.}
  \label{fig:loss_function}
  \end{center}
\end{figure*}

\subsection{Clinical CT volume database for 3D adversarial training}
Deep learning and GAN-based approaches have mostly targeted 2D image slices for artifact reduction. As no clinical database of CT volumes is directly available for learning the characteristics of real metal artifacts, we originally constructed a CT volume database of dental fillings for 3D adversarial training. We corrected 915 clinical CT volumes consisting of head and neck images from the cancer image archive (TCIA) \cite{Clark13} and CT images \cite{Nakao15}\cite{Nakao16} measured from patients who underwent treatment in the Department of Oral and Maxillofacial Surgery of Nara Medical University, Japan. The CT images were scanned on a Siemens SOMATOM Definition AS CT scanner with 120 kV and 200 mAs. This study was performed in accordance with the Declaration of Helsinki and approved by the Institutional Review Board of Nara Medical University Hospital (approval number: 2296). 

To prepare the training data, slice images containing teeth structures from each patient's CT volume were visually checked, and the existence or otherwise of metal artifacts was determined. Artifact-free volumes were classified into image domain $Y$. The CT volumes partly containing metal artifacts were divided into 3D regions with or without metal artifacts. Artifact-free subvolumes were then classified into domain $Y$ and those with metal artifacts were classified into domain $X$. Thus, artifact-free volumes were obtained from both metal-free patient volumes and other patient volumes by excluding the regions with metal artifacts. 

A total of 56 volumes (12 artifact-free CT volumes and 44 volumes with metal artifacts) were randomly selected from the database and used solely as test data for evaluation. The other CT subvolumes consisting of 539 artifact-free volumes (10491 images) and 320 volumes (5655 images) with various patterns of real metal artifacts was used for adversarial training. Each volume consists of 5--43 image slices with $512 \times 512$ pixels. There are no paired data from the corresponding 3D regions that belong to both image domains. Adversarial training for all subsequent experiments was performed using this database under the unsupervised setting. As the proposed MAR targets a wide range of artifacts that appear in soft tissues and bones, [-1000HU, 1000HU] was normalized to [-1, 1]. 

\subsection{3D generative adversarial nets}
The proposed volumetric GAN was designed as an extension to CycleGAN's image-to-image translation framework \cite{Zhu17}. There is a possibility that the 3D distributions of metal artifacts or anatomical structures would not be learned sufficiently well using the conventional 2D framework. We argue that the 3D features learned from unpaired clinical datasets are effective for MAR, and build a 3D generative adversarial net using local CT volumes for adversarial training. As illustrated in Fig. \ref{fig:concept}, our volume-to-volume translation model includes two mapping functions, $G_Y: X \rightarrow Y$ and $G_X: Y \rightarrow X$. Two adversarial discriminators $D_X$ and $D_Y$ are also introduced, where $D_X$ aims to distinguish between volumes ${x}$ and ${G_{X}(y)}$, and $D_Y$ aims to distinguish ${y}$ from ${G_{Y}(x)}$. 

Here, a training sample $x$ or $y$ is an unpaired local volume that consists of $N$ spatially continuous image slices. Fig. \ref{fig:loss_function}(a) shows training samples $x$ and $y$ of clinical CT data in the case of two image slices forming one training unit ($N=2$). The six image sets on the left show the $X \rightarrow Y$ translation, and the right-hand images show $Y \rightarrow X$. It can be confirmed that the local volume contains a spatially continuous distribution of meal artifacts. Fig. \ref{fig:loss_function}(a) is a successful case in which the metal artifacts in the original volume $x$ have been reduced in the translated volume ${G_{Y}(x)}$, and ${G_{X}(y)}$ translated from metal-free sample $y$ include "fake" artifacts represented by the generator $G_X$. 

As there are infinitely many mappings $G$ that induce the input volumes to the target domain, the objective should be designed to fit the mapping functions for the purpose of MAR; the translation should reduce the metal artifacts, recover underlying image features, and preserve native anatomical structures in the original image. The basic objective of CycleGAN can be described as 
\begin{align}
  \label{eq:fullobj}
  \mathcal{L}_{cgan}(G_{X}, G_{Y}, D_{X}, D_{Y}) &= \mathcal{L}_{adv} (G_{Y}, D_{Y}, X, Y) \\ \nonumber 
                                 &+ \mathcal{L}_{adv} (G_{X}, D_{X}, Y, X) \\ \nonumber
                                 &+ \lambda \mathcal{L}_{cyc} (G_{X}, G_{Y}). \nonumber
\end{align}
Here, $\mathcal{L}_{adv}$ refers to adversarial loss, defined as
\begin{align}
  \label{eq:adv}
  \mathcal{L}_{adv} (G_Y, D_Y, X, Y) &= \mathbb{E}_{\bm y}[\log D_{Y}(y)] \\ \nonumber
                                     &+ \mathbb{E}_{\bm x}[\log(1-D_{Y}(G_Y(x)))], \nonumber
\end{align}
where $G_Y$ tries to generate volumes $G_Y(x)$ that are similar to volumes in the target domain $Y$, while $D_Y$ aims to distinguish between $G_Y(x)$ and real samples $y$. Adversarial loss measures the performance of $D$. $\mathcal{L}_{cyc}$ refers to the cycle consistency loss, which is expressed as
\begin{align}
  \label{eq:cyc}
  \mathcal{L}_{cyc}(G_X, G_Y) &= \mathbb{E}_{\bm x} [||G_X(G_Y(x))-x||_{1}] \\ \nonumber
                            &+ \mathbb{E}_{\bm y} [||G_Y(G_X(y))-y||_{1}], \nonumber
\end{align}
where the first term quantifies the reconstruction error between the original image $x$ and the image $G_X(G_Y(x))$ generated through the translation $X \rightarrow Y \rightarrow X$. Similarly, the second term evaluates the cycle consistency of the translation $Y \rightarrow X \rightarrow Y$. The weight $\lambda$ controls the relative strength of the adversarial and cycle consistency loss. $G_{X}$ and $G_{Y}$ are trained such that $\mathcal{L}_{cgan}$ is minimized, and $D_X$ and $D_Y$ is adversarially trained by maximizing $\mathcal{L}_{cgan}$. 

The objective function in (\ref{eq:fullobj}) and its extension can perform successful translations in a variety of medical applications \cite{Liang19, Kida19}. However, it fails to learn the effective translation that targets the metal artifacts, and often modifies native anatomical structures simultaneously. This tendency became significant in the volume-to-volume translations conducted in a preliminary study, and so we explored improved objectives for volumetric MAR.

\begin{figure*}[t]
  \begin{center}
  \includegraphics[width=170mm]{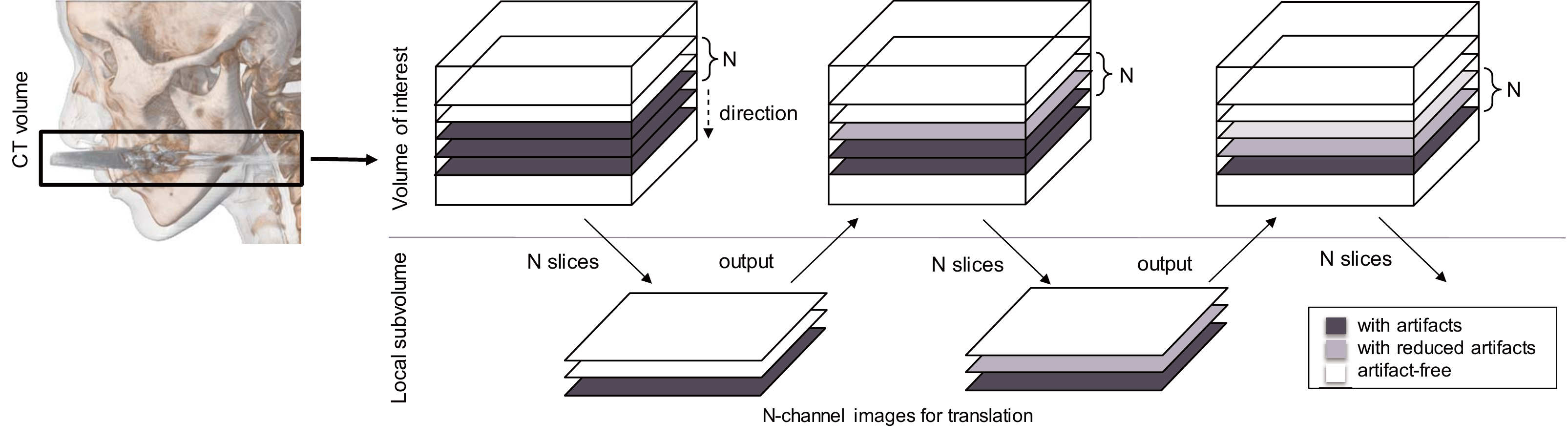}
  \caption{Volume-to-volume translation model for local subvolumes. $N-1$ slices in the next subvolume are replaced by the results modified in the previous translation. This sequential update process aims to reduce the possibility of subvolumes consisting of all low-quality images with strong artifacts.}
  \label{fig:3dgan_flow}
  \end{center}
\end{figure*}

\subsection{Regularized objective function for metal artifact reduction}
To fit the mapping functions targeting metal artifacts for translation, we consider the following two loss functions.\\

\subsubsection{Intensity loss}
As shown by training sample $x$ in Fig. \ref{fig:loss_function}(a), the regions affected by the metal artifacts are often limited or sparse against the entire image space. Although strong artifacts affect a wide range of pixels in a 2D image slice, the 3D distribution is not dense in the volumetric space. We consider these characteristics of the metal artifacts and introduce an intensity loss to reduce the space of possible mapping functions. The intensity loss is defined as a regularization term using the L1 norm, which assigns a penalty to the difference in CT values between the original image $x$ and the translated image $G_Y(x)$. This penalty is expressed as
\begin{align}
  \label{eq:int}
  \mathcal{L}_{int} = \mathbb{E}_{\bm x} ||G_Y(x) - x||_{1} + \mathbb{E}_{\bm y} ||G_X(y) - y||_{1}.
\end{align}

As shown in Fig. \ref{fig:loss_function}(b), this regularization induces an output distribution that is close to an input distribution in CT values, and aims to ensure that the generators translate the sparse artifacts rather than dense anatomical structures. Note that the intensity loss differs from the identity loss \cite{Liang19}\cite{Liao19}, which penalizes $|G_Y(y) - y|$. This loss regularizes the generator, giving an identity map if the sample ${y}$ in the target domain is provided as the input to the generator $G_Y$. The performance of the identity loss for MAR is investigated in the experiments.

\subsubsection{Feature loss}
The distribution of metal artifacts varies in the number of dental fillings, their locations, and other conditions of X-ray propagation. Similar feature patterns, such as dark bands, streaks, and bright areas around metal objects \cite{Gjesteby16}, can be observed. The generator $G_X$ should reduce such artifact-derived features from the original images, whereas the generator $G_Y$ should add similar features to the artifact-free images, as illustrated in Fig. \ref{fig:loss_function}(c). We consider these symmetric characteristics of the translation required for MAR and further constrain the mapping functions by introducing a feature loss. In this study, the feature loss is defined as an L2 norm penalty on the difference in feature space between the values subtracted in the $X \rightarrow Y$ translation and those added in the $Y \rightarrow X$ translation. This can be written as follows:
\begin{align}
  \label{eq:fea}
  \mathcal{L}_{fea} = \mathbb{E}_{\bm x, \bm y} &|| f(x - G_Y(x)) - f(G_X(y) - y ) ||_{2} \\
                    + \mathbb{E}_{\bm x, \bm y} &|| f(G_X(y) - G_Y(G_X(y))) - f(G_X(G_Y(x)) - G_Y(x))||_{2}, \nonumber
\end{align}
where $f$ is the function that calculates the feature values of the input images. This loss induces the generators targeting location-independent, visually similar artifacts for translation. In this study, the pretrained VGG16 network \cite{Simonyan14} is used as a feature encoder, and deep image features are used to evaluate $\mathcal{L}_{fea}$. 

\subsubsection{Full objective}
The full objective $\mathcal{L}$ is defined as 
\begin{align}
  \label{eq:myfullobj}
  \mathcal{L} = \mathcal{L}_{cgan} + \lambda_{int} \mathcal{L}_{int} + \lambda_{fea} \mathcal{L}_{fea},
\end{align}
where $\lambda_{fea}$ and $\lambda_{int}$ are the weights that control the importance of each loss function. The artifact reduction model $G^*_Y$ is obtained by solving  
\begin{align}
  \label{eq:minmax}
  G^*_X, G^*_Y = {\rm arg}\min_{G_X, G_Y}\max_{D_X, D_Y} \mathcal{L}(G_X, G_Y, D_X, D_Y).
\end{align}

To implement the objectives, U-net \cite{Ronneberger15} is employed as the generator and VGG16 \cite{Simonyan14} provides the discriminator. The training volumes were randomly selected from the clinical CT database and applied to the 2D U-net and VGG16 networks as spatially continuous $N$-channel images. These were applied to the developed framework at each epoch to adversarially train $(G_X, D_X)$ and $(G_Y, D_Y)$.

\subsection{Volume-to-volume translation}
Given a patient's whole CT volume with metal artifacts, the trained artifact reduction model $G^*_Y$ translates it to the target domain $Y$. In the preliminary study, we found that relatively moderate or weak artifacts can be effectively reduced, but the correction of strong artifacts with a wide range of missing pixels was case-dependent. To obtain better image quality, we introduce an improved translation model that considers the geometric property of the metal artifacts, as illustrated in Fig. \ref{fig:3dgan_flow}. 

The translation starts from the top or bottom subvolume, and sequentially updates the volume of interest. In the first process, the first subvolume is replaced by the translated output in the original volume of interest. Here, the artifacts contained in the first subvolume are expected to be weak or sparse, as only a few of the $N$ slices are affected by small parts of the dental fillings. In the second process, the next subvolume is translated. This subvolume will overlap with the previous subvolume, and therefore a new slice and modified $N-1$ slices with reduced artifacts are used for the next translation. This sequential update process reduces the possibility of subvolumes consisting entirely of low-quality images with strong artifacts, which can occur in a single update process. Better image quality can be achieved from these sequentially modified subvolumes with moderate or reduced artifacts. The property of this volume-to-volume translation model is quantitatively analyzed in the experiments.

\begin{figure*}[t]
  \centering
  \includegraphics[width=175mm]{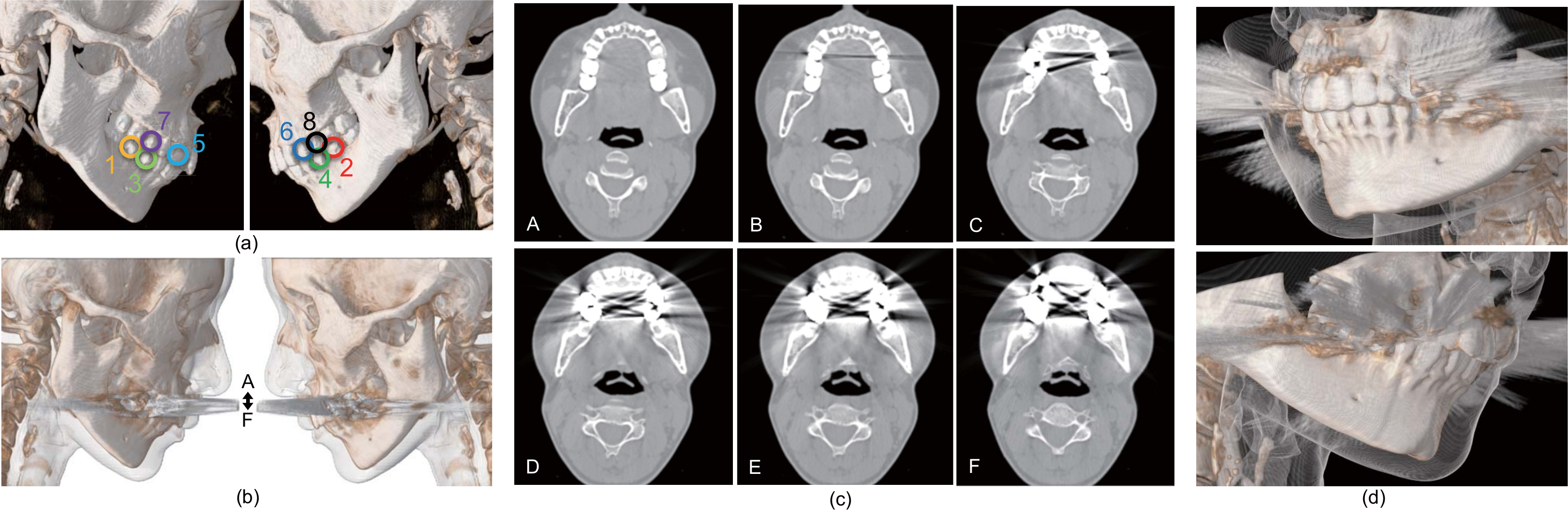}
  \caption{Typical example of the simulated metal artifacts generated from eight dental fillings: (a) selected teeth for modeling metal fillings, (b) generated artifacts and 3D region, (c) 2D image slices, and (d) volume visualization of the simulated metal artifacts. } 
  \label{fig:metal_artifact_database}
\end{figure*}

\section{Experiments}
Three experiments were designed to investigate the performance of the proposed methods: a quantitative comparison, 3D property analysis, and clinical evaluation with expert surgeons. The overall framework was implemented using Python 3.6.8, U-net \cite{Ronneberger15}, VGG16 \cite{Simonyan14}, and the Adam optimizer. A computer with a graphics processing unit (CPU: Intel Core i7-9900X, Memory: 32 GB, GPU: NVIDIA TITAN RTX) was used throughout the experiments. For the regularization parameters, values of $\lambda_{cyc} = 10.0$, $\lambda_{fea} = 1.0$, and $\lambda_{int} = 25.0$ were used after examining several parameter sets. 

\subsection{Metal artifact database}
For a quantitative evaluation of the MAR performance, paired CT volumes (that is, artifact-free volumes as ground truths and corresponding volumes with metal artifacts) are required. To obtain such paired test data, CT-image artifacts were simulated for each metal-free clinical patient volume. To synthesize complex patterns of metal artifacts generated from multiple dental fillings, we manually created volumetric binary labels by extracting 3D regions of eight teeth from the metal-free CT volumes. As shown in Fig. \ref{fig:metal_artifact_database}(a), the first and second were randomly selected from the back teeth. The third and fourth were also extracted from the back teeth close to the first/second teeth. This situation is often seen in real patient data, where two dental fillings adjacent to each other yield strong artifacts due to photon starvation. The fifth and sixth images were selected from the front side teeth, and the other two teeth were randomly chosen from the remaining side teeth. Consequently, eight volumetric metal labels representing 1--8 dental fillings were prepared by combining the selected eight teeth in order. 

The metal artifacts were simulated based on the same procedure and parameters used in \cite{Zhang18}, where metal-inserted volumes were reconstructed using filtered back projection from simulated sinograms. The main differences lie in the volumetric datasets and low-quality images that are simulated from the multiple dental fillings often found in clinical images, especially in elderly patients. Fig. \ref{fig:metal_artifact_database}(c) shows examples of the metal artifacts generated from the volumetric labels with eight virtual dental fillings. Image slices A--F correspond to the 3D region indicated in Fig. \ref{fig:metal_artifact_database}(b). The appearance of the metal artifact changes continuously slice-by-slice, and missing pixels or black bands are generated according to the density of the metal regions. The volumetric distribution of the synthesized artifacts is visualized in Fig. \ref{fig:metal_artifact_database}(d). In this study, a total of 96 volumes containing different patterns of simulated artifacts were created from 12 metal-free CT volumes. The aim of the experiments was to investigate the 3D GAN-based MAR results relating to 3D anatomical structures and complex metal artifacts.

\subsection{Quantitative evaluation}
The first experiment was designed to enable quantitative and qualitative comparisons between the image quality produced by the proposed methods and that given by existing methods. The artifact-free patient volumes were used as references, and the paired metal artifact database with 1--8 virtual dental fillings were used as the original volumes for this experiment. The root mean square error (RMSE) and the structural similarity (SSIM) index \cite{Wang04} were calculated as quantitative error metrics between the reference and the MAR results. SSIM is a good error metric for evaluating the recovery of anatomical structures and the remaining strength of artifacts \cite{Zhang18}\cite{Liao19}. The SSIM index ranges between 0 and 1, with higher values indicating better image quality. We refer to the proposed methods as \textbf{3DGAN}, with a suffix indicating the number of image slices used. For instance, 3DGAN5 means that a local volume with five images was used for volume-to-volume translation. 3DGAN1 means image-to-image translation using on the regularized objective proposed in this paper.

\subsubsection{Baselines}
Liao et al. \cite{Liao19} recently reported extensive evaluation results from a quantitative comparison between their unsupervised MAR (called the artifact disentangle network, ADN) and existing supervised/unsupervised methods using simulated metal artifact images. As ADN produced comparable performance to the supervised methods \cite{Zhang18}\cite{Wang18}, we compared the proposed 3DGAN with the following existing CycleGAN-based methods, including ADN, in the context of an unsupervised learning framework.\\
\textbf{CGAN} \cite{Zhu17} An unsupervised image-to-image translation method using cycle consistency loss for adversarial training.\\
\textbf{CGAN+ID} \cite{Taigman16}\cite{Liang19} This model uses an identity loss that regularizes the generator to be an identity map when images of the target domain are provided as input. \\
\textbf{ADN} \cite{Liao19} Similar to the CGAN+ID model, but with an artifact consistency loss in the image space to constrain the artifact difference. 

\begin{figure}[t]
  \centering
  \includegraphics[width=88mm]{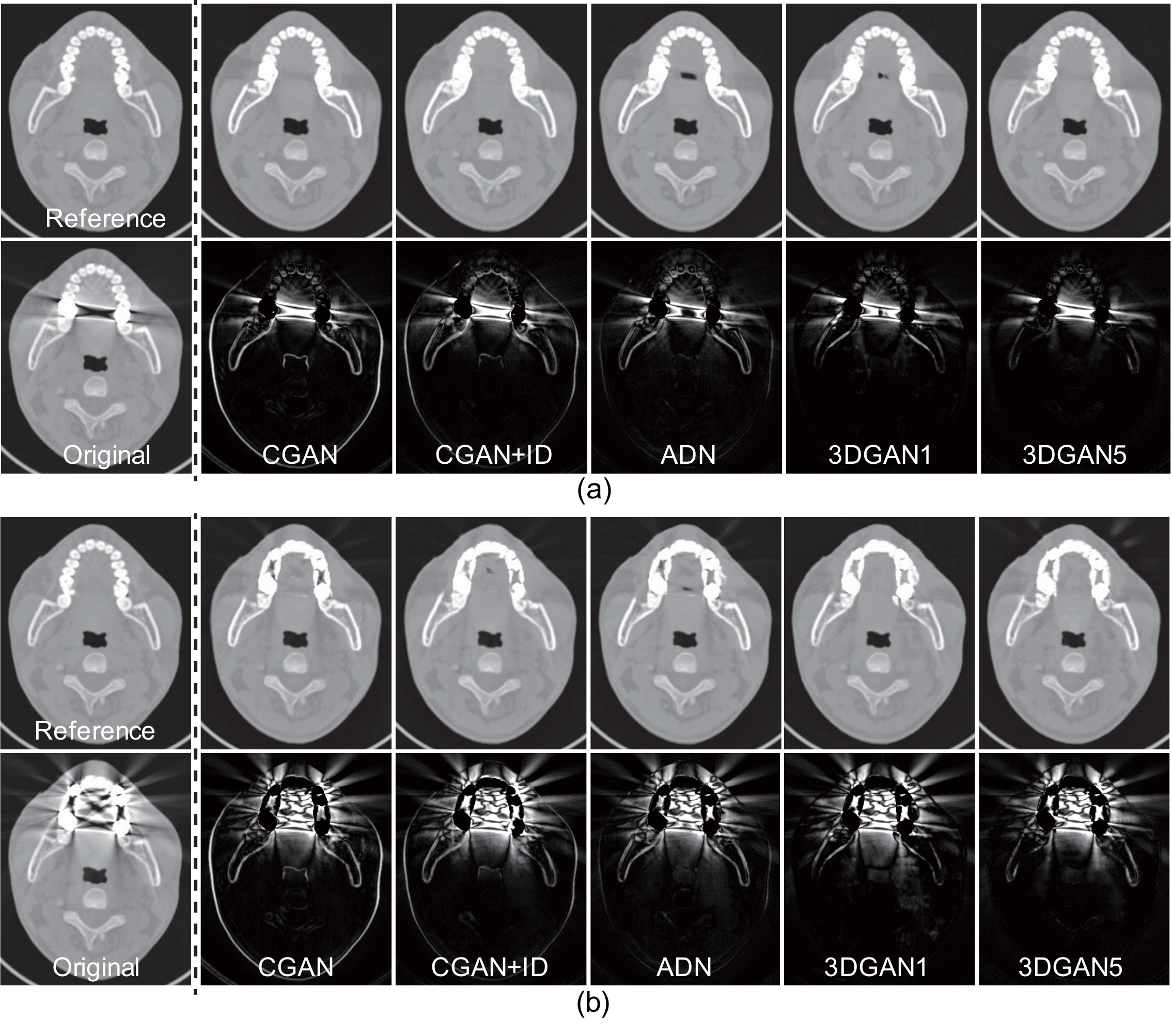}
  \caption{Corrected results of the simulated metal artifacts generated from (a) three and (b) eight metals. The upper images are the reference (ground truth) and the results obtained using CGAN, CGAN+ID, ADN, 3DGAN1, and 3DGAN5. The lower images are the original image and the difference between the corrected results and the reference.}
  \label{fig:results_sim_2d}
\end{figure}

\begin{figure*}[t]
  \centering

  \includegraphics[width=180mm]{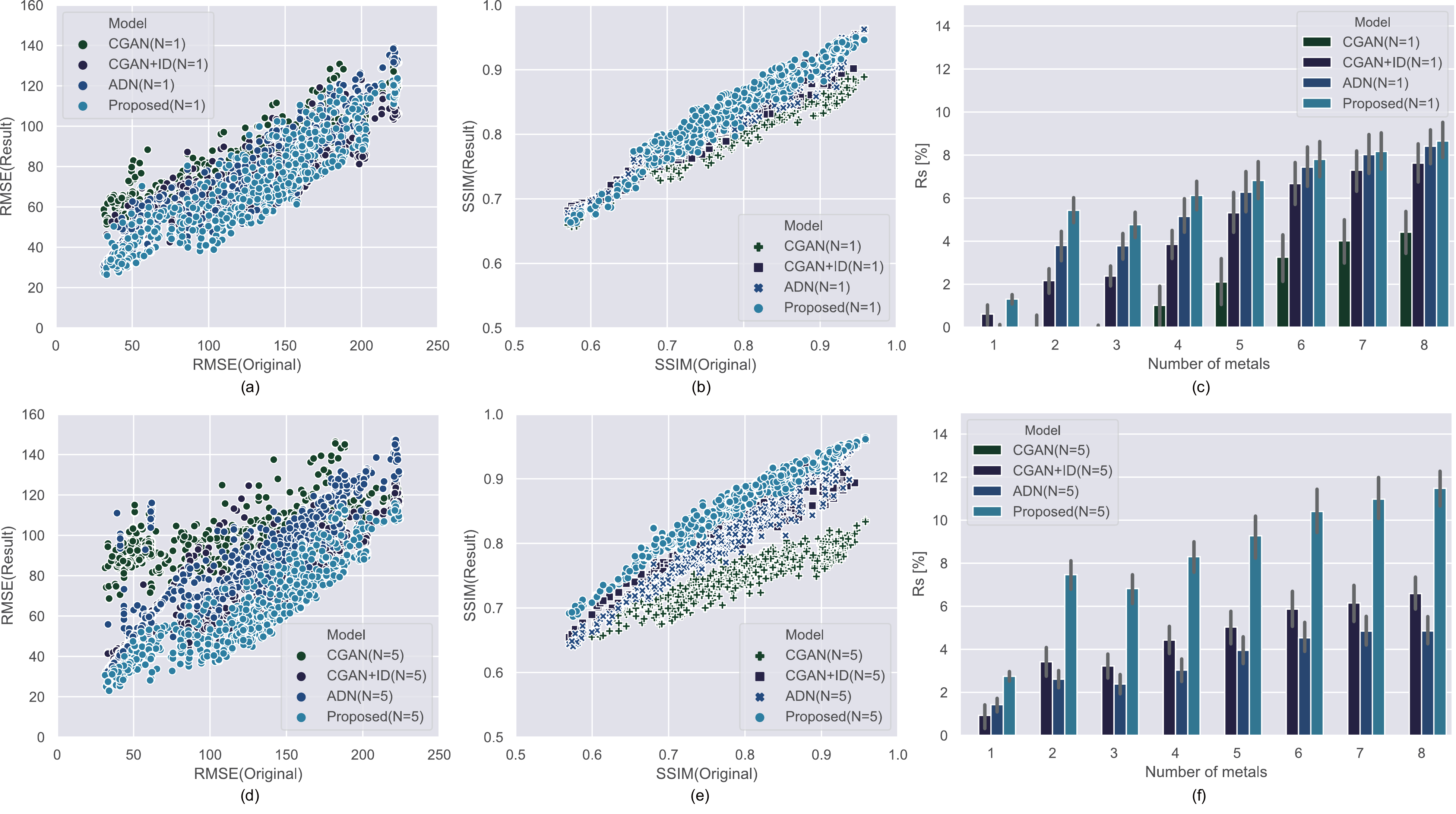}
  \caption{Quantitative comparison of the corrected images with respect to various metal artifact patterns. 2D plots of (a) RMSEs and (b) SSIMs of original and corrected results, and (c) performance improvement $R_{s}$ in 2D translation. 2D plots of (d) RMSEs, (e) SSIMs and (f) $R_{s}$ in 3D translation.}
  \label{fig:scatter_plots}

\end{figure*}

\begin{figure*}[t]
  \begin{center}
 \includegraphics[width=180mm]{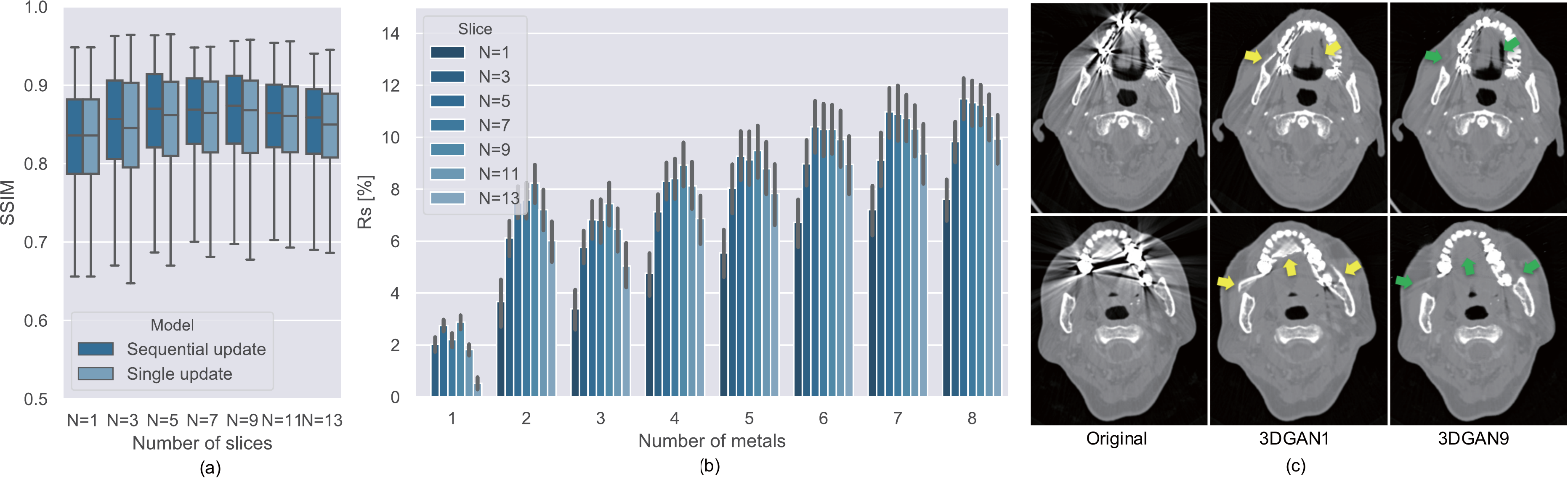}
  \caption{MAR properties of 3D adversarial training: (a) SSIMs of the obtained results in different 3D settings, (b) improvement rate with respect to the number of metals, and (c) corrected images in 2D and 3D translation. }
  \label{fig:3dgan_property}
  \end{center}
\end{figure*}

\subsubsection{Comparison against baselines}
Fig. \ref{fig:results_sim_2d} shows the corrected results for the simulated artifacts generated from three and eight metals. The upper images are the reference (ground truth) and the results obtained using CGAN, CGAN+ID, ADN, 3DGAN1, and 3DGAN5. The lower images are the original image and the absolute difference between the corrected results and the reference. Although the artifacts have been corrected in the results of CGAN and CGAN+ID, the subtraction images show that the edges of the soft tissues and mandibular structures were also modified. This could yield inadequate deformation of the anatomical shape. ADN and 3DGAN1 preserve the CT values of soft tissues; however, some teeth and mandibular structures were wrongly corrected, and residual artifacts remain in Fig. \ref{fig:results_sim_2d}(a). 3DGAN5 achieved better image quality against different artifact patterns while preserving anatomical structures.   

Table \ref{table:quantitative_comparison} lists the median values of RMSE and SSIM for the original and corrected images with respect to the reference images. The results for artifacts generated from different numbers of metals ($m= 1$, $4$, or $7$) are listed. 3DGAN1 achieved slightly superior performance over the baseline values, and 3DGAN5 outperformed the other methods. There is a relatively large difference between the values given by 3DGAN5 and 3DGAN1, which may imply that volume-to-volume translation is robust to various patterns of the real artifacts. To further analyze the performance, the relationship between the SSIMs of the original images and the corrected results were investigated by introducing an error metric, the improvement rate of image quality $R_{s}$. This is defined as
\begin{eqnarray}
R_{s} = \frac{SSIM_{corrected} - SSIM_{original}}{SSIM_{original}} \times 100
\end{eqnarray}
This index takes higher values when stronger artifacts are corrected and the reference CT values are adequately recovered. Additionally, we applied the loss functions of the baselines to volume-to-volume translation and compared the resulting performance with our 3DGAN5. The baselines were originally developed for 2D images. Their 3D extension is worth investigating to clarify the loss function design. 

Figs. \ref{fig:scatter_plots}(a) and \ref{fig:scatter_plots}(b) plot the RMSE and SSIM values of the original images and the corrected results, respectively. Fig. \ref{fig:scatter_plots}(c) shows the $R_{s}$ values of the four image-to-image translation models with respect to the number of metals $m$. The performance of 3DGAN1 is slightly better when the CT volume has fewer than five metals. Figs. \ref{fig:scatter_plots}(d) and \ref{fig:scatter_plots}(e) plot the RMSE and SSIM values of the volume-to-volume translation models, and Fig. \ref{fig:scatter_plots}(f) shows their $R_{s}$ values. The difference in performance between 3DGAN and the baselines becomes significantly larger. Specifically, 3DGAN5 achieved better image quality than its 2D model, whereas the performance of the baselines became worse in the 3D setting. These results suggest that the loss function of the baselines wrongly corrects soft tissues and teeth structures, and appropriate regularization is required for volume-to-volume translation with higher-dimensional inputs. The regularization terms of the proposed method could contribute to proper image correction that targets metal artifacts while preserving anatomical structures.

\begin{table}[t]
  \begin{center}
      \caption{Median values of RSME and SSIM for original and corrected images with respect to reference images.}
  \scalebox{0.75}{      
    \begin{tabular}{ccccccc}
    \hline
      $m=1$& \textbf{Original} & \textbf{CGAN} & \textbf{CGAN+ID} & \textbf{ADN} & \textbf{3DGAN1} & \textbf{3DGAN5} \\ \hline
      \textbf{RSME} & 48.5 & 63.4 & 50.9 & 46.8 & 41.1 & 36.7 \\
      \textbf{SSIM} & 0.854 & 0.915 & 0.901 & 0.924  & 0.920  & 0.937 \\ \hline
    \end{tabular}
  } 

  \vspace{3mm}

  \scalebox{0.75}{      
    \begin{tabular}{ccccccc}
      \hline
      $m=4$& \textbf{Original} & \textbf{CGAN} & \textbf{CGAN+ID} & \textbf{ADN} & \textbf{3DGAN1} & \textbf{3DGAN5} \\ \hline
      \textbf{RSME} & 127.3 & 74.7 & 64.6 & 67.4 & 59.9 & 54.4 \\
      \textbf{SSIM} & 0.801 & 0.816 & 0.836 & 0.850  & 0.859  & 0.875 \\ \hline
    \end{tabular}
  }

  \vspace{3mm}

  \scalebox{0.75}{      
    \begin{tabular}{ccccccc}
      \hline
      $m=7$& \textbf{Original} & \textbf{CGAN} & \textbf{CGAN+ID} & \textbf{ADN} & \textbf{3DGAN1} & \textbf{3DGAN5} \\ \hline
      \textbf{RSME} & 156.4 & 89.2 & 79.2 & 82.2 & 81.3 & 75.6 \\
      \textbf{SSIM} & 0.752 & 0.784 & 0.803 & 0.813  & 0.817  & 0.836 \\ \hline 
      \end{tabular} 
  }

  \label{table:quantitative_comparison}
  \end{center}
\end{table}

\subsection{3D property analysis}
Subsequent experiments were designed to confirm the characteristics of volume-to-volume translation based on 3D adversarial training. Limitations on loading the multi-channel images into the GPU memory meant that we investigated the cases of $N=1, 3, 5, 7, 9, 11, 13$ used for the local input volume of 3DGAN. Fig. \ref{fig:3dgan_property}(a) shows box plots of the SSIMs obtained by different 3DGAN settings (number of slices and update model). Fig. \ref{fig:3dgan_property}(b) summarizes the improved image quality with respect to the number of metals $m$. The box plots show that the proposed sequential update model outperforms the single update model in volume-to-volume translation. The mean value of the SSIM increased for $N \leq 9$, and 3DGAN9 achieved the best SSIM and improved image quality. 3DGAN11 and 3DGAN13 produced worse performance, specifically in the cases of few metals. Fig. \ref{fig:3dgan_property}(c) shows two examples of visually different results obtained by 2D and 3D translation for clinical CT volumes. As 3DGAN1 learns the image-to-image translation for the CT image slices, it tends to mistranslate artifact regions and generate unnatural correction results around the teeth, tongue, and mandible, as illustrated by the yellow arrows. In contrast, 3DGAN9 provides satisfactory correction results (green arrows) while reflecting 3D anatomical structures, even for voxels that are difficult to recover. 

\begin{figure*}[t]
  \begin{center}
 \includegraphics[width=170mm]{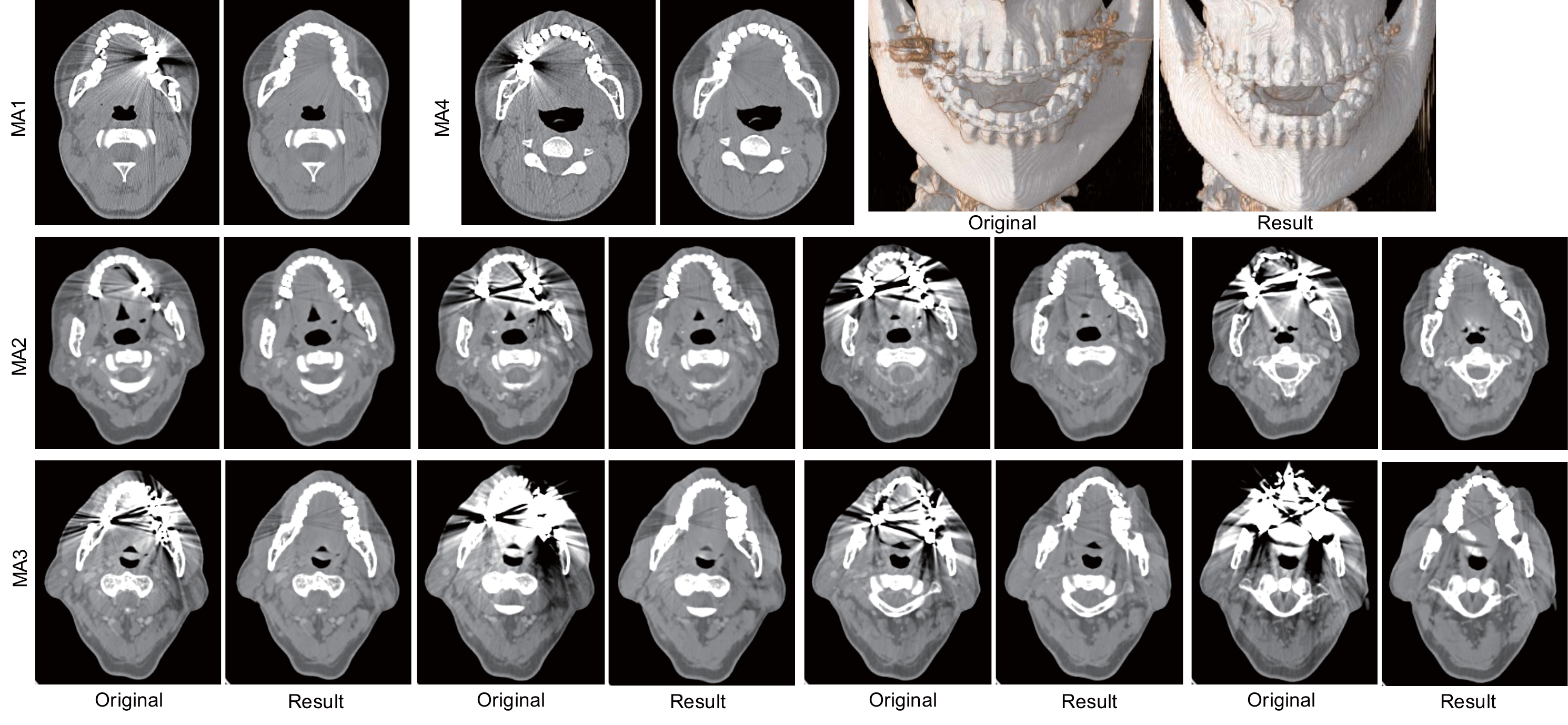}
 \caption{Corrected clinical CT images with a variety of artifact patterns. MA2 and MA3 contain strong artifacts with a wide range of dark bands and scattered missing pixels. Visually satisfactory recovery of missing voxels could be achieved despite the lower quality of the original images.}
 \label{fig:result_real}
  \end{center}
\end{figure*}

\subsection{Clinical evaluation}
To confirm the image quality of the proposed MAR method in clinical applications, subjective evaluations with expert oral and maxillofacial surgeons were conducted. Three oral surgeons (senior specialists with 15, 26, and 36 years of experience, respectively) participated in this experiment. They selected four CT volumes (MA1, 2, 3, and 4) as clinically typical examples with different artifact patterns. 

\subsubsection{Artifact reduction results for clinical images}
Fig. \ref{fig:result_real} shows the corrected results of the four volumes. 3DGAN9 was used for artifact reduction. MA1 contains typical metal artifacts from a single dental filling, and the CT values and textures of the soft tissues have been recovered. MA2 and MA3 contain strong artifacts with a wide range of dark bands and scattered missing pixels, and the three original images show continuous changes in the artifact patterns of the local volumes. Although residual artifacts and deformation of soft tissues or teeth structures can be seen, visually satisfactory recovery of missing information, i.e., 3D structures and CT values, was achieved despite the lower quality of the original images. MA4 contains orthodontic appliances and was measured with the mouth open for diagnostic use. The results show that 3DGAN removed most parts of the appliances and associated metal artifacts while preserving 3D teeth structures. This example confirms the robustness and 3D property of the proposed 3DGAN, as such cases were not included in the training datasets. 

\begin{table*} [t]
  \begin{center}
      \caption{Clinical validation results: average (minimum and maximum) values of quality of artifact reduction (QOAR) and structural accuracy (SA), and duration required for image correction. }
  \begin{tabular}{cccccccccc}
  \hline
  \multirow{2}{*}{} & \multicolumn{2}{c}{QOAR} & & \multicolumn{2}{c}{SA} & & \multicolumn{2}{c}{Duration}    \\ \cline{2-3} \cline{5-6} \cline{8-9}
                         & Manual & Proposed & & Manual & Proposed & & Manual [min] & Proposed [sec] \\ \hline
  MA1 & 3.3 (3 -- 4) & 4.0 (4 -- 4) & & 3.3 (3 -- 4) & 4.0 (4 -- 4)  & & 12 & 6.6  \\
  MA2 & 2.0 (2 -- 2) & 3.0 (3 -- 3) & & 2.0 (3 -- 3) & 3.0 (3 -- 3)  & & 50 & 11.9 \\
  MA3 & 2.0 (2 -- 2) & 3.0 (3 -- 3) & & 2.3 (2 -- 3) & 2.7 (2 -- 3)  & & 52 & 13.2  \\
  MA4 & 3.7 (3 -- 4) & 4.0 (4 -- 4) & & 3.3 (3 -- 4) & 4.0 (4 -- 4)  & & 18 & 15.8 \\ \hline
  Avg. & 2.8 & 3.5 & & 2.8 & 3.5 & & 33 & 11.9 \\ \hline
  \end{tabular}
  \label{table:clinical_validation}
  \end{center}

\end{table*}

\begin{figure*}[t]
  \begin{center}
 \includegraphics[width=150mm]{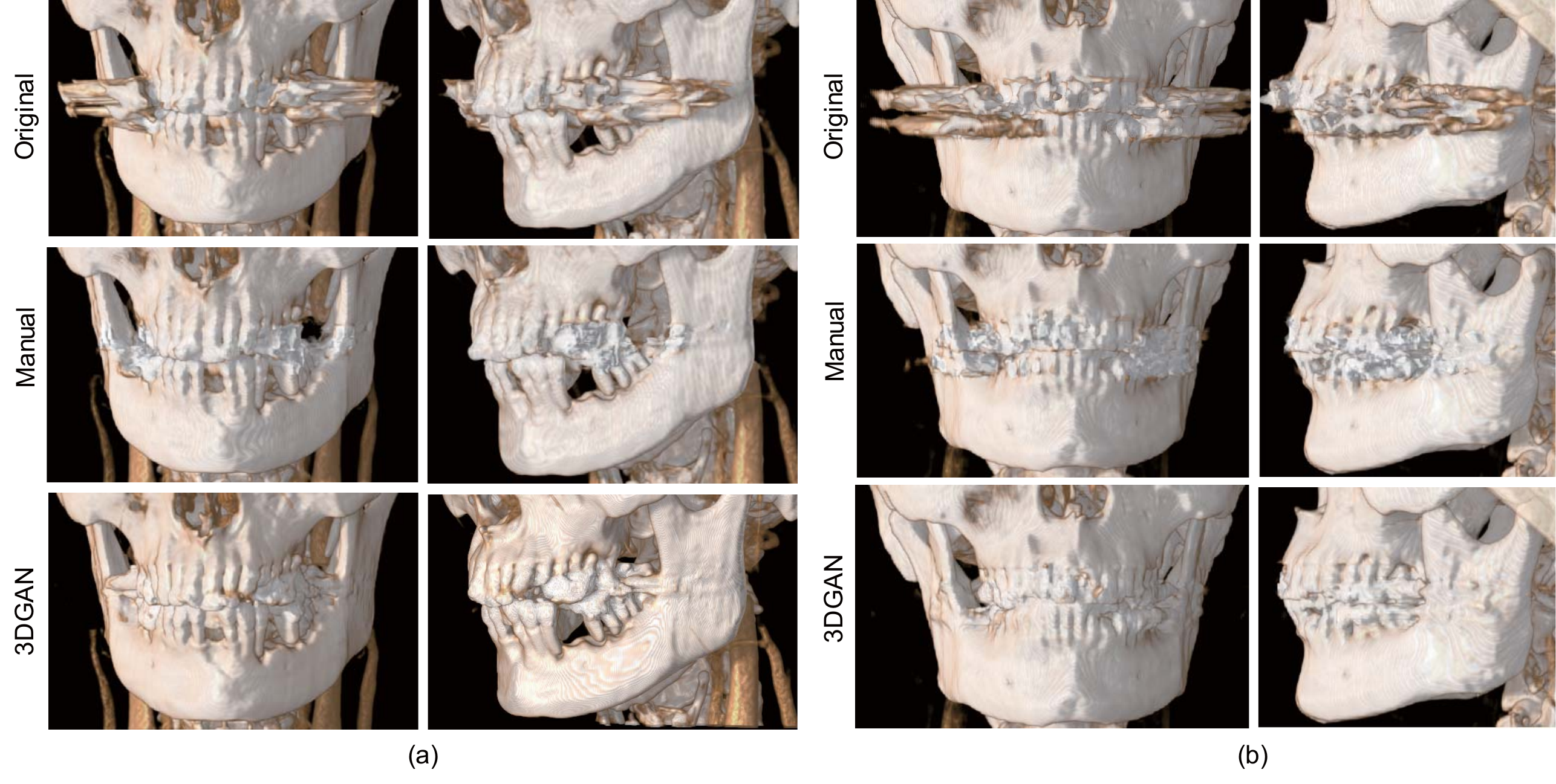}
  \caption{Volume rendered images obtained by manual correction and 3DGAN-based MAR assuming clinical use in surgical planning: (a) MA2 and (b) MA3. 3DGAN generates visually plausible corrections of the 3D teeth structures, whereas manual correction produces an irregular appearance of the teeth surfaces. }
  \label{fig:manual_vs_3dgan}
  \end{center}
\end{figure*}

\subsubsection{Evaluation by expert surgeons}
We considered the clinical use of MAR for surgical planning, specifically in mandibular reconstructive surgery \cite{Nakao15}\cite{Nakao17}, and compared the image quality of two MAR approaches: manual correction by a dental technician (with over 20 years of experience) as the current clinical protocol, and 3DGAN-based MAR. Manual correction is widely employed for strong artifacts with missing pixels, which cannot be recovered by conventional MAR functions implemented in commercial CT devices. (For instance, radiation oncologists also manually correct images for radiation dose planning in radiotherapy applications.) 

The results of the manual and 3DGAN-based MAR were displayed in random order. Each participant checked the volume-rendered image sets of the front, side, and bottom views obtained from the two results and compared their image quality. To evaluate the clinical availability of the results, we defined three evaluation criteria: quality of artifact reduction (QOAR), structural accuracy (SA), and duration of the overall image correction process. QOAR scores the degree of metal artifacts for clinical use, i.e., whether metal artifacts were adequately corrected. SA scores anatomical correctness, i.e., whether the structure of the mandible and the teeth were accurately represented. Both metrics were assigned one of four grades: Excellent: 4, Good: 3, Fair: 2, and Poor: 1. Good was defined as a level with clinically sufficient quality for preoperative planning, and Fair was defined as having a few problems, but with acceptable quality.

Fig. \ref{fig:manual_vs_3dgan} shows volume-rendered images of the 3DGAN-based MAR results and those manually corrected by the dental technician. Table \ref{table:clinical_validation} summarizes the scores obtained from the three surgeons and the duration required for the overall process. The proposed MAR scored higher than the manual correction results in terms of both QOAR and SA for all data, with an average of 3.5. Specifically, all participants considered the MA1 and MA4 results containing moderate artifacts to be excellent. Although MA2 and MA3 included strong artifacts, such as multiple dark bands and scattering noise, the scores show that 3DGAN achieves clinically sufficient image quality and outperforms manual correction by the dental technician. The comments from the surgeons suggest that the metal artifacts have been visually corrected for both images. 3DGAN generates visually plausible corrections of the 3D teeth structures, whereas manual correction produces an irregular appearance of the teeth surfaces. The average duration of image correction required for the two approaches was 11.9 s for 3DGAN and 33 min for manual correction. These results show that the developed framework rapidly provides corrected results for the real CT volumes with strong artifacts and contributes to improved productivity in preoperative or radiotherapy planning. 

\section{Discussion}
To the best of our knowledge, this study is the first to build 3D adversarial nets with a regularized loss function for metal artifact reduction derived from multiple dental metal fillings. The experimental results have shown that 3D adversarial training from unpaired patient CT datasets and volume-to-volume translation can achieve clinically acceptable MAR for clinical images with strong artifacts. To clarify the focus of this research, we have concentrated on quantitative comparison with unsupervised learning methods. For a quantitative comparison between supervised MAR using convolutional neural networks (CNNs) and state-of-the-art MAR methods \cite{Meyer10}\cite{Stayman12}, refer to \cite{Gjesteby17} and \cite{Zhang18}. Reference \cite{Liao19} compares CycleGAN-based MAR and existing supervised/unsupervised methods.

To date, most artifact reduction approaches, including deep learning studies, have considered artifact reduction in images with few metal objects \cite{Zhang18}\cite{Liao19}. Supervised or unsupervised learning based on synthesized images requires elaborate simulation of complex and abundant variations of artifact patterns, and the difference from real artifacts becomes problematic. There are no well-trained CNN models for the strong, complex artifacts that often appear in clinical CT volumes generated from multiple (for instance, more than four) dental fillings. The experiments using the metal artifact database showed that the proposed 3DGAN directly learned from unpaired patient images has robust artifact reduction ability, even for simulated artifacts not included in the training data.

The results of 3D property analysis showed that 3DGAN achieves better image quality than 2D translation; however, the performance was slightly worse when using 11 or 13 image slices. This may be because the increased number of slices reduces the number of volumes available for adversarial training, or because the dimension of the input volumes might be greater than required to learn 3D anatomical features, resulting in overfitting. To further improve the proposed method, the range of the Hounsfield units could be optimized for the focused regions. The exploration of adversarial training designs targeting specific artifacts and anatomical structures and semi-supervised learning for MAR are interesting topics for future studies.

\section{Conclusion}
This paper introduced MAR methods based on unsupervised volume-to-volume translation learned from clinical CT images. The results of experiments using 915 CT volumes from real patients demonstrated that the proposed 3DGAN has an outstanding capacity to reduce strong artifacts and to recover underlying missing voxels while preserving the 3D features of soft tissues and tooth structures. Unsupervised learning directly from unpaired clinical images has potential applications to various artifact patterns that are difficult to handle using filter-based or prior knowledge-based MAR approaches. Future challenges include improving the adversarial training and prediction framework and the application to other clinical fields, specifically radiotherapy planning.

\bibliographystyle{IEEEtran}

\ifCLASSOPTIONcaptionsoff
  \newpage
\fi

\vfill
  
\end{document}